\documentclass[sigconf,natbib=false]{acmart}
\AtBeginDocument{%
  }

\RequirePackage[
  datamodel=acmdatamodel,
  style=acmnumeric,
  ]{biblatex}

\setcopyright{acmlicensed}
\copyrightyear{2025}
\acmYear{2025}
\acmDOI{XXXXXXX.XXXXXXX}
\acmConference[CUG25]{Computing Horizons: CUG 2025}{May 04--08,
  2025}{New Jersey}
\acmISBN{978-1-4503-XXXX-X/2025/05}

\usepackage{balance}

\begin{document}

\title{Alps, a versatile research infrastructure}

\author{Maxime Martinasso}
\authornote{This work reflects the collective contributions of the CSCS team, who have been essential to the development of the Alps research infrastructure.}
\author{Mark Klein}
\author{Thomas C. Schulthess}
\affiliation{%
  \institution{ETH Zurich, Swiss National Supercomputing Centre (CSCS)}
  \city{Lugano}
  \country{Switzerland}
}

\renewcommand{\shortauthors}{Martinasso et al.}

\begin{abstract}
The Swiss National Supercomputing Centre (CSCS) has a long-standing tradition of delivering top-tier high-performance computing systems, exemplified by the Piz Daint supercomputer. However, the increasing diversity of scientific needs has exposed limitations in traditional vertically integrated HPC architectures, which often lack flexibility and composability. To address these challenges, CSCS developed Alps, a next-generation HPC infrastructure designed with a transformative principle: resources operate as independent endpoints within a high-speed network. This architecture enables the creation of independent tenant-specific and platform-specific services, tailored to diverse scientific requirements.  

Alps incorporates heterogeneous hardware, including CPUs and GPUs, interconnected by a high-performance Slingshot network, and offers a modular storage system. A key innovation is the versatile software-defined cluster (vCluster) technology, which bridges cloud and HPC paradigms. By abstracting infrastructure, service management, and user environments into distinct layers, vClusters allow for customized platforms that support diverse workloads. Current platforms on Alps serve various scientific domains, including numerical weather prediction, and AI research.  

\end{abstract}

\ccsdesc[300]{Computing methodologies~Massively parallel and high-performance simulations}
\ccsdesc[300]{Computer systems organization~Interconnection architectures}
\ccsdesc[100]{Computer systems organization~Multicore architectures}
\ccsdesc[500]{Software and its engineering~Cloud computing}
\ccsdesc[500]{Software and its engineering~Agile software development}

\keywords{HPC, Cloud layers, research infrastructure, platforms}
\begin{teaserfigure}
\includegraphics[width=\textwidth]{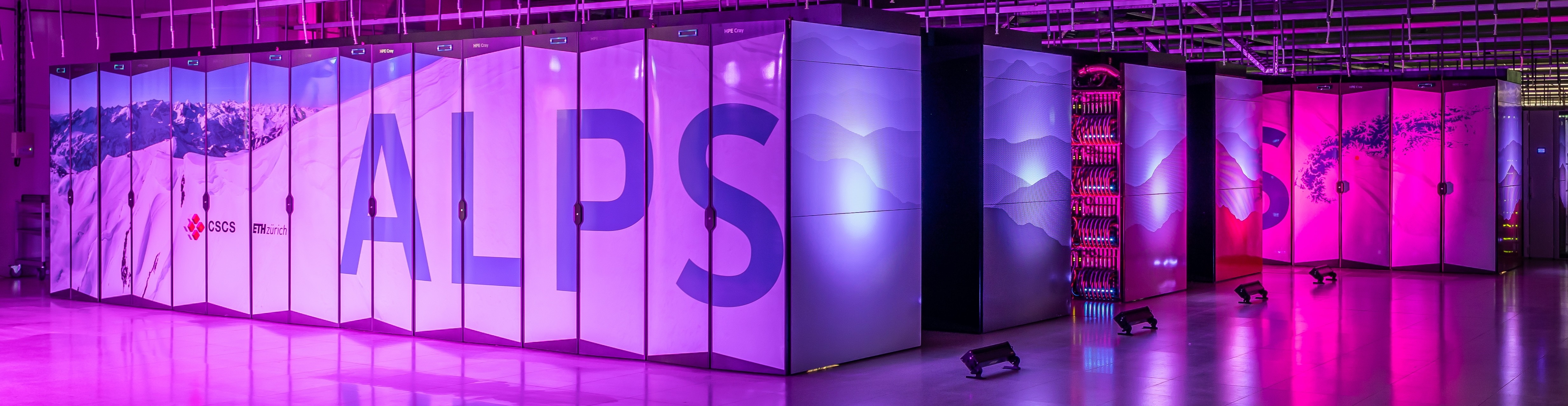}
  \caption{The Alps research infrastructure, 2024.}
  \Description{Picture of a the supercomputer Alps.}
  \label{fig:teaser}
  \Description{A picture of the Alps research infrastructure at CSCS.}
\end{teaserfigure}


\maketitle

\section{Introduction}

CSCS has long been committed to delivering top-ranked HPC systems such as Piz Daint to support its scientific community as part of its mission. However, traditional management of HPC systems has presented significant challenges. Issues such as system upgrades, varying software stacks, and diverse storage requirements have tested the capability of systems like Piz Daint to provide flexibility and separation of concerns among set of services.

A large shared HPC system is provided to various scientific groups, each with unique requirements, including specific workflows and software stacks. In such a shared system that offers a single set of services to all scientific communities, the service offerings often limit scientists' ability to customize them to meet their needs. Consequently, scientists must adapt their workflows and software stacks to fit the constraints of the system, which hampers scientific efficiency and, ultimately, discovery.

To address these challenges, CSCS adopted a new approach when designing its next-generation infrastructure, Alps~\cite{Alps}. Architecturally, Alps is built on a simple yet transformative principle: every resource operates as an endpoint within a global high-speed network. These resources are organized into independent clusters. Using the cloud model and layer abstraction, custom services are deployed on these clusters, enabling the creation of independent tenant and platform services tailored to specific needs. These clusters, known as versatile software-defined clusters (vClusters)~\cite{vCluster1}~\cite{vCluster2}, are defined by versioned, human-readable recipes and instantiated by pipelines, ensuring immutability once deployed.

This flexibility in providing HPC capabilities to scientists increases the number of services to manage and operate compared to a single offering. Even if services are similar, such as a Slurm batch scheduler, the software versions and configurations vary across the numerous vClusters deployed on Alps. This new service model also necessitates a novel approach to operations by an HPC data center like CSCS. Existing concepts, such as service reliability engineering and end-to-end responsibility, must be adapted to accommodate both the constraints of HPC technology and the culture at the HPC center.

This paper presents the design of Alps, the concept of vClusters, and their operational model challenges. By co-designing services together with the scientific communities, Alps is not merely a successor to Piz Daint but a versatile research infrastructure tailored to a wide range of scientific domains.

\section{Limitations of HPC systems}

CSCS has operated systems like Piz Daint for several decades. During this time, the engineering teams have faced recurrent issues that have hindered CSCS's ability to provide optimal services and opportunities to its user community. The main issues are summarized in the following subsections.

\subsection{Vertically integrated stack}

HPC systems are designed as vertically integrated solutions that combine hardware and software to optimize the performance of scientific applications. These systems rely on highly tuned software stacks, such as numerical libraries and low-level network libraries, tailored to the hardware infrastructure. These software packages are bundled together in large images provided by the vendor.

Although this vertical integration optimizes system performance, it imposes significant constraints on users and scientists, especially during updates. When updates occur, multiple components, such as compilers and libraries, are changed simultaneously forcing users to rebuild applications and do extensive re-validations, which is both costly and time-consuming. Furthermore, identifying the source of any problems that arise can be complex and can result in conflicts of responsibility among support groups. In addition, developing scientific applications on vendor-specific stacks presents challenges in portability. Containers offer a workaround for user-defined stacks, but come with complex configurations that require specialized container engines~\cite{sarus} and are not suitable for all use cases, in particular the one of developing and debugging applications. 

\subsection{Monolithic service offering}

The vertically integrated stack includes a set of services and the programming environment. Although adding additional services is possible, it is complicated and requires adaptation to the integrated stack. For instance, introducing a new Linux distribution with its own kernel is nearly impossible, and even less extreme cases, such as implementing a non-vendor-supported batch scheduler or a custom programming environment, can be challenging. Even simpler tasks, like changing the version of Slurm, are constrained by the vendor's release schedule and expected version. This traditional, vendor-provided-or-nothing approach to service management sacrifices flexibility and composability, limiting the ability to offer diverse HPC ecosystems on shared infrastructure. This inefficiency hampers serving multipurpose use cases that require diverse set of services, such as programmatic APIs, high-throughput computing, or interactive access, leading to suboptimal resource utilization.

\subsection{Operational constraints}

Another limitation of this traditional approach is the operational constraints it imposes. When issues, bugs, and security vulnerabilities are discovered, the vendor cannot provide immediate fixes that can be easily deployed. Instead, fixes are released in large bundles on a monthly or quarterly cadence, resulting in significant delays. During this period, workarounds are employed, increasing the heterogeneity of the operational mode and complicating bookkeeping. Once an update is released, it undergoes intensive regression testing~\cite{reframe} by the center to ensure that key applications and workflows are not affected. Often, they are, leading to further delays in applying the update. Additionally, the update process requires complete downtime of the machine for several hours or days, reducing its usage and limiting the number of acceptable updates.

\section{The Alps infrastructure}

\subsection{Objectives}

The mission of CSCS is to develop and operate a high-performance computing and data research infrastructure that supports world-class science in Switzerland. To fulfill this mission, CSCS deploys HPC systems such as Piz Daint, which are offered to a community of researchers. Resource access is open and managed through a transparent, peer-review process, establishing the offered resource as a research infrastructure. The inception of Alps stemmed from the need to provide a new HPC infrastructure to replace Piz Daint.

Over the past decades, the cloud business and technology model has been introduced to provide custom services on generic IT infrastructure, primarily using virtualization. While this model is appealing for scientific computational use cases, its application to HPC infrastructure has proven challenging due to the inability to use virtualization to ensure high performance. CSCS engineering teams recognized that integrating certain aspects of cloud technology into HPC infrastructure is feasible by adopting a service-oriented infrastructure and grouping network-attached resources into independent clusters. Applying this cloud technology reduces the limitation of HPC systems. This realization led to the development of the vCluster technology together with the deployment of the Alps infrastructure.

The resulting objectives of the Alps infrastructure are designed to provide greater flexibility and composability, enabling more efficient use of the infrastructure, supporting diverse user needs, and allowing incremental service upgrades for individual vClusters. These services and solutions, crafted in collaboration with the scientific communities themselves, position the Alps infrastructure as a versatile research infrastructure catering to diverse scientific disciplines.

\subsection{Hardware resources}

Alps is built as a heterogeneous infrastructure, featuring a mix of CPUs and GPUs interconnected via a high-speed network. Specifically, Alps is an HPE Cray EX system that leverages Slingshot for efficient communication. Table~\ref{tab:alps_hardware} presents the types of nodes currently available. Alps is the first HPC system in the world to deploy GH200 chipsets on a large scale.
Alps is equipped with multiple storage systems to support diverse workloads. These include a 100PB ClusterStor HDD system and a 3PB ClusterStor SSD system, both utilizing the Lustre file system for high-performance data access. Additionally, a 1PB VAST storage system is available, offering additional flexibility and performance for specific use cases.

The GH200 nodes in Alps achieved a measured HPL performance of 434 PFlops, earning the system 7th place in the November 2024 Top500 ranking.

\begin{table}
\caption{Current Node Composition of Alps.}
\label{tab:alps_hardware}
\begin{tabular}{lcc}
\toprule
\textbf{Node Type} & \textbf{Number of Nodes} & \textbf{Total GPUs} \\ \midrule
AMD Rome-7742          & 1,024  & -      \\ \midrule
NVIDIA A100            & 144    & 576    \\ \midrule
AMD MI250x             & 24     & 96     \\ \midrule
AMD MI300A             & 128    & 512    \\ \midrule
Grace-Hopper (GH200)   & 2,688  & 10,752 \\
\bottomrule
\end{tabular}
\end{table}

\subsection{Platform objectives}

Alps is managed by the Cray System Management (CSM)~\cite{CSM} software, referred to as the management plane. CSM consists of a set of microservices with APIs that enable node configuration and management, such as rebooting and image deployment. A key aspect of CSM is its capability to label and manage metadata of compute nodes, facilitating the management of groups of nodes. Combined with the Alps network, based on the Slingshot network~\cite{slingshot}, which enables resource segregation, it becomes possible to manage groups of resources individually within clusters. These clusters are defined by a set of labeled resources and a set of services deployed onto those resources. The clusters are then grouped into platforms offered to specific communities.
The service management of platforms key objectives for Alps include: 
\begin{itemize} 
\item Ensuring the provision of independent platforms, combining HPC resources and customizable services, offered to scientific communities or tenants. 
\item Guaranteeing scalable, reliable, and efficient management of multiple platforms to reduce engineering costs. 
\end{itemize}

To enable the management of multiple independent platforms and clusters, CSCS has engineered the versatile software-defined cluster or vCluster~\cite{vCluster1}~\cite{vCluster2} using cloud concepts applied to HPC.

\section{vCluster and vService concepts}
The concept of a vCluster within HPC infrastructure is introduced as a key technology bridging the HPC and cloud paradigms. While cloud technology enables the deployment of custom services on virtualized infrastructure, vClusters offer similar abstractions without the need for virtualization, ensuring efficient use of HPC resources while providing service flexibility. 
The vCluster model employs a three-tier abstraction layer: Infrastructure as Code, Service Management, and User Environments. This provides a separation of concerns between the HPC resources provided at the infrastructure level and the services deployed on top of them in the service management layer. Finally, adaptable services and user environments enable multiple diverse scientific communities to operate their own platforms offering customization and services specialized for their use cases. A platform can deploy multiple vClusters, each defining a specific resource configuration and the services it hosts. Such services deployed on a vCluster are referred as vServices.

\begin{figure}[ht]
\centering
\includegraphics[width=\linewidth]{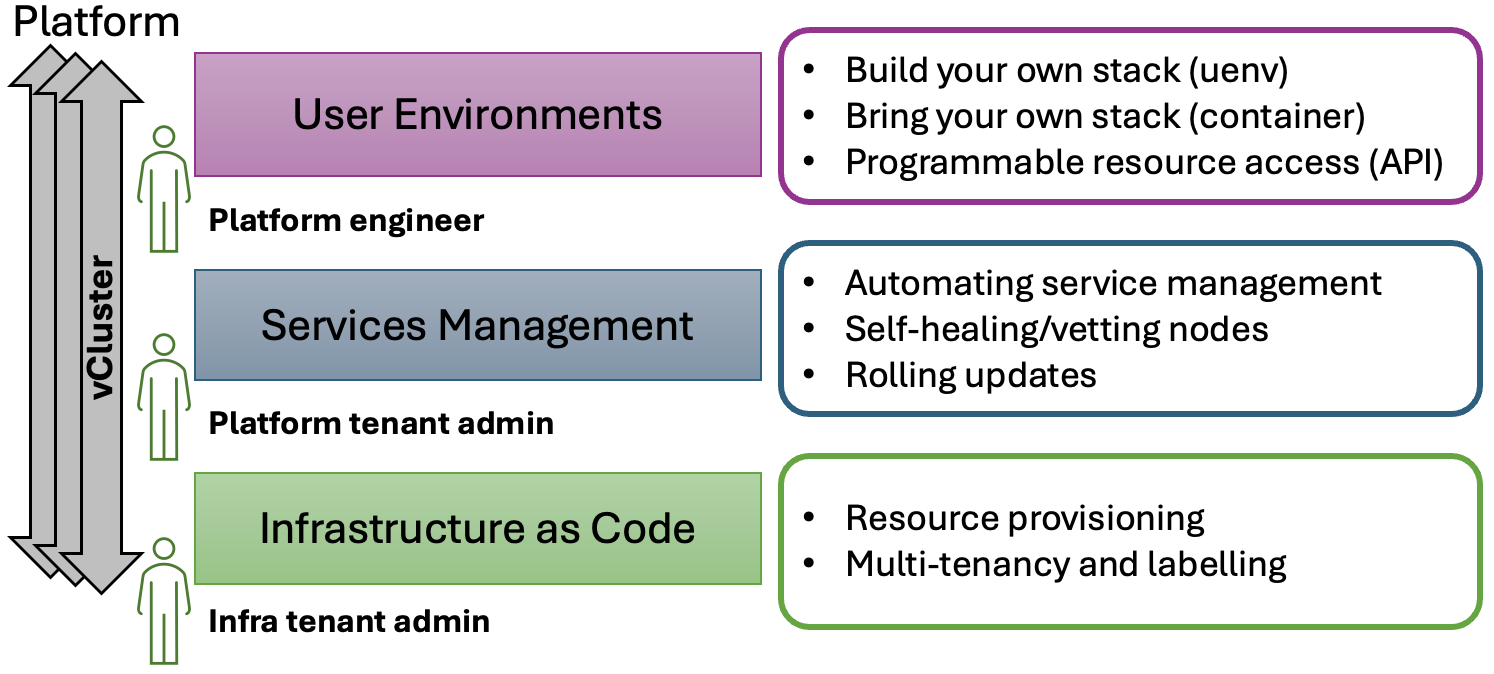}
\caption{The vCluster technology employs a three-layer abstraction that separates resource provisioning, service deployment mechanisms, and the selection of required services for scientific platforms, ensuring flexibility and modularity in HPC environments. A platform is composed by one or several vClusters.}
\label{vCluster3layer}
\Description{Diagram of the 3 layers: infrastructure as code, service management and user environments}
\end{figure}

To ensure consistency and reliability, the vCluster leverages software pipelines and role-based access control, with configurations defined in textual form, executed through pipelines, and remaining immutable once deployed.
Control over these layers is distributed among various roles, including infrastructure administrators, platform tenant administrators, platform engineers, service managers, and scientists.
Designed to be infrastructure-independent and vendor-neutral, vCluster technology offers adaptability across various providers.
Figure~\ref{vCluster3layer} presents the three-layer concept of the vCluster technology.

\subsection{Layer 1: Infrastructure as Code}

The goal of this layer is to provision and manage HPC resources in a multi-tenancy environment and deploy minimal base images, which are further configured by the service management layer.

\subsubsection{System management interface:}
HPC systems rely on a configuration system, often referred to as the management or control plane, to handle essential tasks such as rebooting nodes, monitoring system health, and deploying images. Examples include the HPE Cray System Management (CSM)~\cite{CSM} software, and tools like OpenStack for HPC~\cite{openstack4hpc} or OpenCHAMI~\cite{openchami}. For provisioning resources to vClusters, it is crucial to associate label to compute nodes to enable grouping. Such grouping allows independent resource management at the system level and provides a list of resources used by the upper service management layer to deploy the vCluster services. Several management system like CSM and OpenCHAMI provide a micro-service handling this task such as rebooting all nodes associated with a specific label.

To achieve vendor-agnostic flexibility, the infrastructure layer must abstract the complexities of HPC configuration and resource management from the underlying control plane. The tool Manta~\cite{Manta} has been developed to provide such an abstraction and interface. Currently, Manta interfaces with CSM and OpenCHAMI. Manta offers several key features, including multi-site management and role-based access control (RBAC). Multi-site management allows a single Manta instance to oversee distinct and distributed infrastructures, while RBAC ensures that administrators can only manage their assigned groups of resources. Further integration of Manta using Terraform~\cite{terraform} and Google's cluster toolkit~\cite{GoogleHPC} is ongoing, aiming to enable the provisioning of identical vClusters to the cloud.

\subsubsection{Compute node image and network:}
Once the resources are grouped, the nodes can be booted using a base image. This deployed node image is minimal, containing only essential components such as kernels, drivers, and hardware libraries, often provided by vendors to support specialized hardware configurations. Resource isolation is achieved when necessary through network segregation using technologies such as VLANs~\cite{vlan} and PKEYs~\cite{pkey}. High-speed networks like Slingshot and InfiniBand are capable of creating network isolation through PKEYs, VLANs and enforcing them at the switch level ensuring network isolation of the booted nodes.

\subsubsection{Multi-tenant storage:}
Traditional Lustre-based HPC storage offerings face multi-tenancy challenges. However, many (and new) vendors are developing software-defined HPC storage solutions that support multi-tenancy. For instance, VAST, DDN, and Weka offer solutions with these features. Nevertheless, the integration of their solutions into a complete high-performance ecosystem remains to be assessed, and their performance compared to a Lustre file system needs further evaluation. Alternatively, current approaches include mounting distinct directories of a large Lustre filesystem for each vCluster within a platform or dedicating storage systems to specific tenants or platforms.

Overall, the infrastructure technology layer abstracts resource provisioning and management, assigns resources based on tenant-specific labels, and segregates them via the network layer (when needed). The labeled resources are then provided to the upper service management layer to deploy services.

\subsection{Layer 2: Service management}

The service management layer within a vCluster plays a critical role in managing and orchestrating platform services, referred to as vServices, and ensuring a cohesive HPC experience. It provides tools to automate the management and deployment of vServices, such as user identity, resource access, workload scheduling, security, user environments, container engines, interactive access, and many other specific services. This layer also manages the health of compute nodes and, in case of minor issues, can trigger automatic reboots using Manta. In the event of infrastructure problems, it can set aside nodes for investigation. Note that a node joining an existing vCluster is automatically configured with the vServices, thanks to the vService orchestrator detailed in this Subsection.


\subsubsection{vCluster and vService definition:}

\begin{figure}[ht]
\centering
\includegraphics[width=\linewidth]{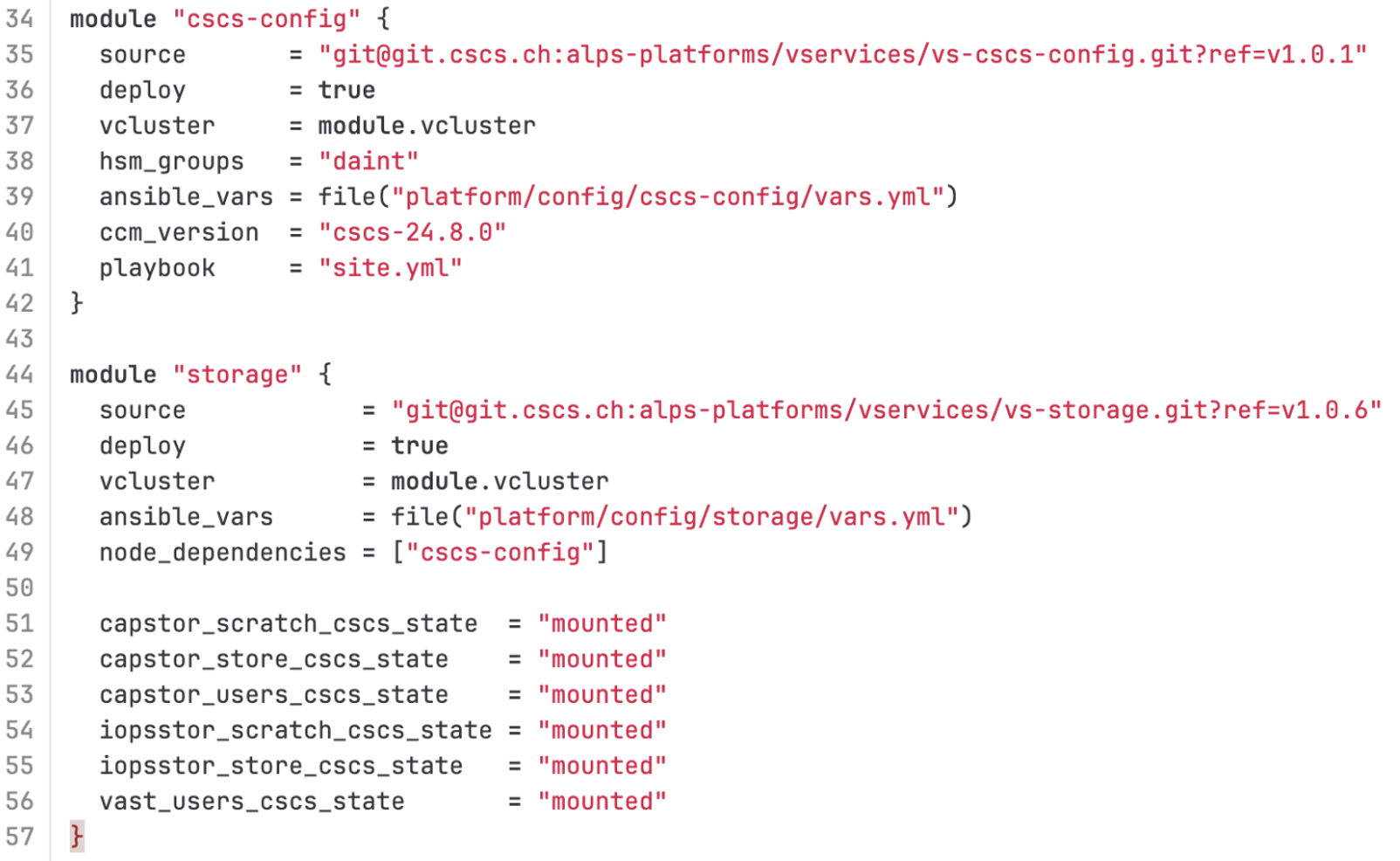}
\caption{Example of the manifest for the vCluster Daint. The first vService \emph{vs-cscs-config} in the manifest defines the common configuration of the base image specific to CSCS. It also sets the vCluster name \emph{daint} using the \emph{hsm\_groups} attribute which matches the label used to group resources in the infrastructure layer. The second vService \emph{vs-storage} specifies which storage is mounted on each node. Note the Git reference and version number in the \emph{source} attribute, which provide the location of the vService code for deployment and the associated version to use for that vCluster.}
\label{vClusterManifest}
\Description{A text file that shows a vCluster manifest composed of vService references.}
\end{figure}

A vCluster is defined by using declarative text files managed in a Git repository which includes a manifest of vServices, along with a list of configuration files associated with each vService. Each item in the manifest refers to a Git repository and a version number, containing the code necessary to deploy the vService. This approach allows vServices to be developed independently and released in different versions, enabling vClusters to select the appropriate vService and version number for deployment. Figure~\ref{vClusterManifest} presents an example of the definition of a vService within a vCluster manifest. A vService repository includes the code and tests needed to execute, install, and remove software packages or configurations, as well as to start and stop processes on each compute node within the vCluster. During the past two years, the CSCS engineering teams have generated approximately 900 merge requests across the Git repositories of vServices. Currently, 16 vClusters operate under this model, with each vCluster manifest containing around 100 lines of code, corresponding to about 10-20 vServices.

\subsubsection{vCluster deployment mechanism:}

A vCluster deployment consists of two main components. The first component is the resource plane, which includes the HPC compute nodes and a vService orchestrator responsible for managing the vService lifecycle~\cite{CloudNativeHPC}. In our case, this orchestrator is Nomad~\cite{nomad}, which automates tasks such as installing and removing software packages in the base image, applying and cleaning configurations, and starting or stopping daemons like Slurm daemons. The second component is the service plane, composed of service nodes where a Kubernetes orchestrator is permanently deployed. The service plane is used for starting processes that are not needed on the compute nodes but required by the vServices, such as a Slurm control daemon. Unlike the resource plane, the service plane is not part of the high-speed network but remains connected to it.

\begin{figure}[ht]
\centering
\includegraphics[width=\linewidth]{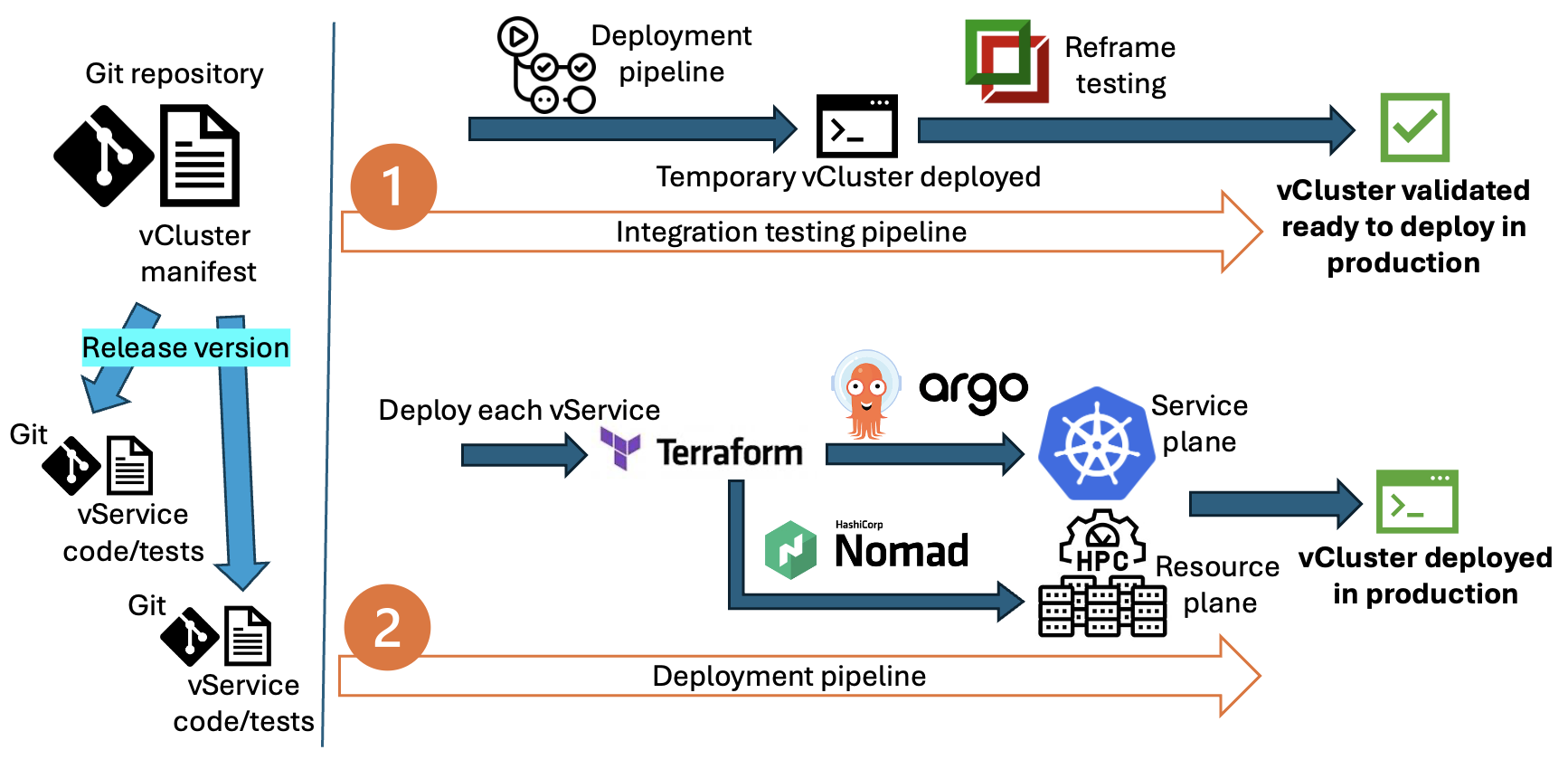}
\caption{Visual representation of the integration testing and deployment pipelines.}
\label{vClusterDeploy}
\Description{A picture showing pipelines and elements used by them.}
\end{figure}

The vCluster deployment process begins by retrieving the node list from the infrastructure layer using Manta~\cite{Manta}, based on the vCluster label. The vService orchestrator clients are then deployed on these nodes using Cloud-init~\cite{cloud_init}. If a node joins or leaves the vCluster, the necessary setup is applied or removed accordingly. Next, a Kubernetes cluster is instantiated by using ArgoCD~\cite{argocd} on the service plane to handle vService processes outside of the compute nodes.
The main step in the deployment is the instantiation of all vServices that constitute the vCluster. To ensure automation and consistency, pipelines are used for both vCluster and vService deployment and are presented in Figure~\ref{vClusterDeploy}.

\begin{enumerate}
    \item 
Integration pipeline: Executes a deployment pipeline -- detailed in point (2) -- to create a temporary, small-scale version of the vCluster on a few nodes. Runs Reframe~\cite{reframe} to perform vService and vCluster integration tests. If the tests pass, the temporary vCluster is destroyed and a production pipeline can be executed.
\item Production deployment pipeline: Triggers the deployment of vService elements across both the service and resource planes. Informs both orchestrators to restart or manage the lifecycle of updated or new vServices.
\end{enumerate}

The deployment pipeline uses Terraform~\cite{terraform} as a common interface for deployment across both planes.
If a vCluster manifest is modified after a reviewed merge request, an operator can trigger the integration testing pipeline. Upon successful completion, the deployment pipeline can be initiated to update the vCluster services. This process can utilize rolling updates; however, for certain vServices, such as a batch scheduler (primarily Slurm), extra care must be taken. This includes considerations like moving compute nodes from the old version to the new version of the vService.

\subsubsection{Self-healing/vetting of compute nodes:}

A large resource plane consisting of thousands of nodes and tens of thousands of GPUs is prone to component failures, leading to node failures. These failures can occur during job execution. In this scenarios, the service management layer must ensure that the vCluster is resilient to these failures. A specialized vService is deployed to monitor node health at the start and end of each compute job with elements both in the resource and service planes. If a node fails, this vService will detect it and execute a pipeline to attempt automatic recovery by rebooting the node. If recovery is not successful or not possible, this vService will remove the node from the vCluster by executing actions on both the batch scheduler and the vService orchestrator. The infrastructure tenant administrator will be informed of the repair request along with logs. Once the node is repaired, another pipeline will reintegrate it into the vCluster.

In general, the service management layers provide the necessary tools and orchestrators to enable the deployment and life-cycle management of vServices. It ensures immutability of deployment by using automated actions inside pipelines.

\subsection{Layer 3: User environments}

User environments comprise the selected platform services designed to support the scientific community utilizing that platform. These services include programming environments, programmatic and interactive access, batch scheduling (e.g., Slurm), container orchestration (e.g., Kubernetes), identity and access management, and various other tools. Each service operates independently and is deployed via the second layer of the vCluster architecture.
Table~\ref{vServiceList} presents the catalog of the most common vServices and their functionalities.

\begin{table}
\caption{List of the most common vServices and their functionalities.}
\label{vServiceList}
\begin{tabular}{lll}
\toprule
\textbf{vService} & \textbf{Functionality} \\ \midrule
CSCS config & Applies specific CSCS configuration of Alps  \\ \midrule
Uenv            & Deploys uenv tool     \\ \midrule
Enroot       & Deploys enroot container engine      \\ \midrule
Podman            & Deploys podman container engine   \\ \midrule
Pyxis   & Deploys Slurm plugin for containers  \\ \midrule
CDI & Injects HPC devices within containers \\ \midrule
FirecREST & Sets up login nodes to accept API requests \\ \midrule
IAM & Sets up of SSSD for SSH access \\ \midrule
Network & Sets up DNS and public IPs \\ \midrule
Node validator & Installs Reframe and runs health checks \\ \midrule
Slurm & Deploys Slurm, databases and rest service \\ \midrule
Storage & Mounts and unmounts file systems \\ \midrule
Alpernetes & Connects Alps compute nodes to Kubernetes \\
\bottomrule
\end{tabular}
\end{table}

\subsubsection{Build your own stack:}
Programming environments, however, represent a unique case due to their reliance on an integrated software stack. To enhance flexibility, a tool named uenv~\cite{uenv} has been developed. This technology packages programming environments or fully built scientific applications into single deployable files using the SquashFS format. With role-based access control, platform engineers or users can customize their tools and libraries through a descriptive, text-based configuration. A pipeline leveraging Spack as a dependency manager generates the specified environment or builds the specific application.

At runtime, users can select their desired environment, utilizing integrated tools like modules to set up the appropriate execution context. This approach empowers users with autonomy, allowing them to define and manage their environments independently or provide their own application. It prevents disruptions caused by unexpected administrative changes, eliminating the need for frequent rebuilding and revalidation of applications.

\subsubsection{Bring your own stack:}
Container technology is the primary method for users to bring their own software stack, which is particularly relevant for the AI and ML community. Specific container engines for HPC have been developed~\cite{containersHPC} to ensure seamless integration into HPC ecosystems and to provide access to high-performance devices available on the nodes. The capability to deploy a container on HPC resources involves three main components:
\begin{enumerate}
\item Integration with the batch scheduler to define the container image and environment. This is achieved using a vService that deploys the Pyxis spank plugin. Pyxis parses a TOML file that describes the container requirements.
\item A container engine to start the container by interacting with kernel system calls. Two vServices are available: one for Podman and one for Enroot.
\item A vService that installs a set of plugins enabling access to host devices during container deployment. This uses the OCI standard to remain independent of the selected container engine~\cite{madonna}.
\end{enumerate}

\subsubsection{Programmatic resource access:}
Enabling programmatic access to resources—such as submitting jobs or transferring data in and out of an HPC data center—is a key capability for integrating into scientific ecosystem of tools and practices. FirecREST~\cite{FirecREST} was developed to provide this functionality. It is a web-facing REST API that offers a set of endpoints for interacting with batch schedulers and data management systems.

Beyond being just a REST API, FirecREST provides a complete ecosystem that includes a web gateway for external access, IAM configuration to enable client authentication using credentials, and an SSH connector for interacting with HPC components like Slurm. To ensure seamless integration of vClusters into the FirecREST ecosystem, a vService has been created.

Once FirecREST is set up, it unlocks additional capabilities, such as Jupyter Notebooks for interactive computing, custom web interfaces, workflow engine integration, including AiiDA and Airflow, and connectivity to external CI/CD pipelines. In fact, implementing a FirecREST backend for a tool is relatively simple since FirecREST provides python bindings. As FirecREST is a web-facing interface, all these tools can be deployed by scientific communities either on-premises or in the cloud.

The three layers presented in this architecture incorporate technologies designed to manage vClusters and platforms. However, they also introduce a larger number of components and a more complex management model, effectively redefining the nature of operational activities in the system. 

\section{Evolution of operation}

The term "operation" refers to the set of actions required to keep a system fully functional in a production environment—i.e., capable of reliably handling production workloads. Traditionally, an HPC center operates a limited number of production systems, typically consisting of one large, shared HPC system and a few smaller auxiliary systems tailored to specific use cases.

With Alps, however, the introduction of vCluster technology significantly increases operational complexity. First, the system is now composed of numerous platforms, making it impractical to have single teams to manage each platform individually—doing so would dramatically increase the required number of system engineers. Second, vClusters are immutable by design: they cannot be modified manually but must instead be updated through Git commits, code reviews, and automated pipeline executions. This enforces a shift in operational practices from traditional system administration to software engineering-oriented workflows.

To accommodate this paradigm shift, CSCS has undertaken a comprehensive redesign of its operational model, resulting in a new organizational structure and a redefinition of roles and responsibilities.

\subsection{Organizational structure}

Before the introduction of Alps, CSCS was organized along functional lines, with teams dedicated to broad business objectives. For example, managing Piz Daint primarily involved two large teams: a user support team and an HPC operations team. The user support team was responsible for addressing user requirements and use cases such as resolving tickets related to the programming environment, and ensuring a smooth user experience on the system. The HPC operations team focused on lifecycle management tasks and maintaining the infrastructure’s readiness for production workloads. Each team consisted of roughly 20 engineers.

\begin{figure}[ht]
\centering
\includegraphics[width=\linewidth]{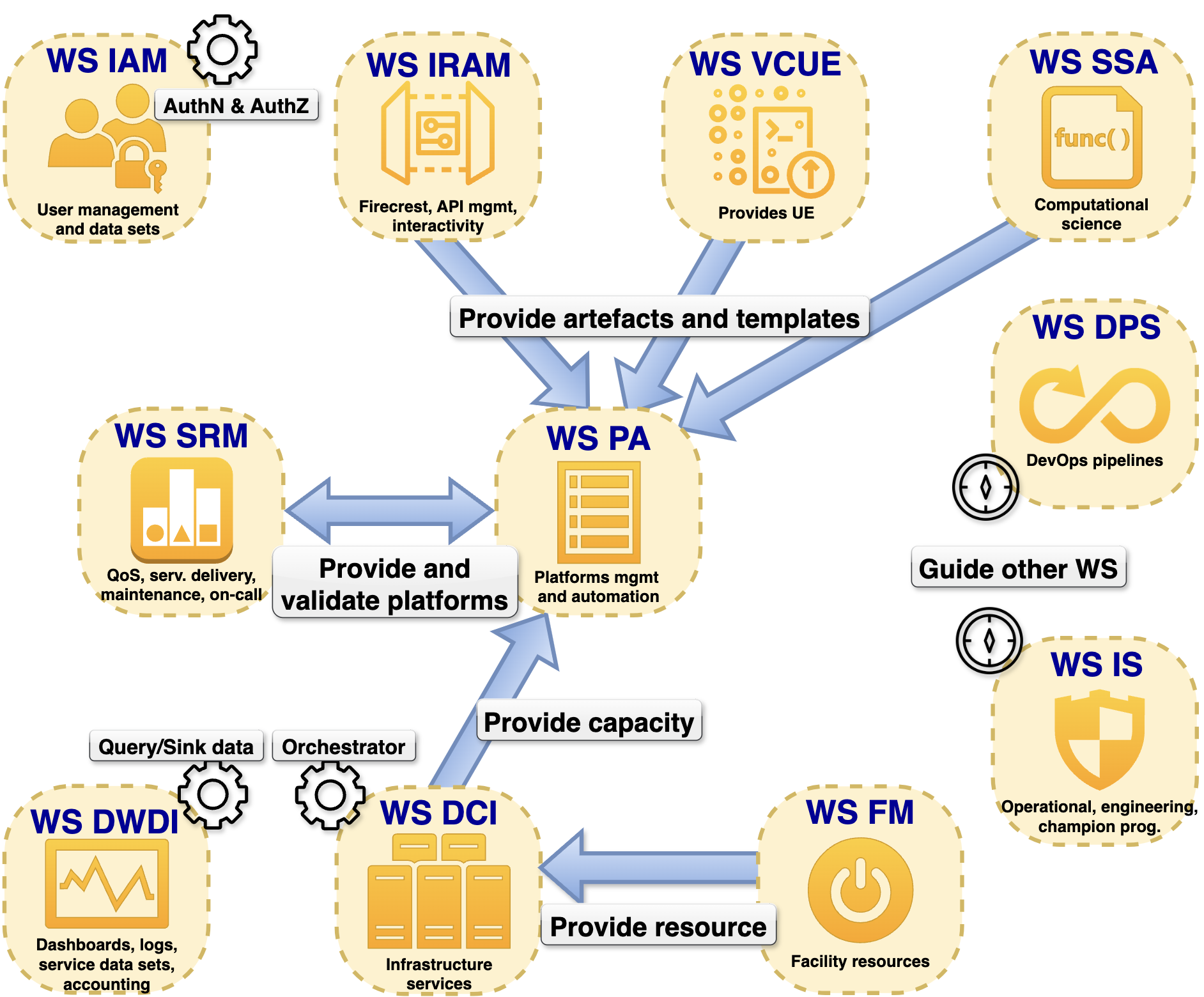}
\caption{The engineering quadrant consists of 11 teams, comprising approximately 100 engineers in total. WS-PA delivers platform automation capabilities to manage the vClusters, playing a central role in the engineering effort alongside WS-SRM, which ensures the quality and service delivery of the platforms. Other teams provide various vServices, such as WS-VCUE for user environments, WS-IRAM for APIs and interactivity, and WS-SSA for scientific code optimized for HPC. WS-DCI offers infrastructure capacity, while WS-FM supplies essential facility resources like power and cooling. Additionally, several teams provide global capabilities and services to the entire engineering quadrant, including WS-IAM for authentication and authorization, WS-DWDI for monitoring and analytics, WS-IS for security, and WS-DPS for DevOps tooling and pipeline management.}
\label{EngQua}
\Description{A picture showing the current organization of the engineering quadrant in 11 teams.}
\end{figure}

With the transition to a service-oriented architecture for Alps, CSCS embarked on a multi-year effort to transform its internal organizational structure. Today, CSCS operates with smaller, dynamic teams, each dedicated to delivering technical solutions within a specific domain. Those teams are named Working Structures (WS). For example, one WS focuses on identity and access management, another develops platform automation tooling, a third manages the infrastructure layer, including CSM, networking, and storage, and another ensures service quality and acts as a liaison with users. In total, CSCS currently operates with 11 distinct WS, although this number has varied over the year, peaking at 14 WS. Figure~\ref{EngQua} presents the current WS and their relationships.

These teams are all part of what is now known as the Engineering Quadrant, which collectively manages the various vClusters and platforms. In addition to the Engineering Quadrant, CSCS has established three other organizational quadrants: Strategy and Governance, Research Infrastructure and Partner Development, and internal Corporate Services.

\subsection{End-to-end responsibility}

In parallel with the reorganization of CSCS, the concept of operations had to evolve to align with the service-oriented architecture introduced by Alps. Unlike the previous structure, where specific teams were dedicated mainly to operational tasks, this new model eliminates standalone operations teams in favor of cross-functional WS that take end-to-end responsibility for service delivery.

Each WS operates under a dual mandate:
\begin{itemize}
    \item 
Solution Development: This includes building the specific technologies of the WS objectives and enabling them using automation, such as embedding new capabilities/software into vServices with developing their deployment pipelines, and designing processes that foster collaboration and consistency across WS.
    \item 
Service Ownership: Since WS teams provide services to both platform users and other WS, they are fully accountable for the quality of the services they deliver. This makes each WS responsible not only for development but also for the operational excellence of their solutions.
\end{itemize}

For example, when a user submits a support ticket, an automated system analyzes the ticket content and routes it to the appropriate WS queue. The assigned WS is then responsible for handling the ticket and delivering a resolution. WS-SRM (Service Reliability Management) oversees this process to ensure overall service quality.

Working structures are therefore empowered and operate with a high degree of independence, owning both the development and operational quality of their services. To maintain alignment across these autonomous teams, a dedicated WS within the Strategy and Governance Quadrant ensures architectural cohesion, while another WS in the same quadrant manages an agile development process to harmonize local WS outcomes with the broader organizational goals.

\subsection{Agile development process}

The quarterly planning process has been implemented to foster coordination and cohesion within the engineering quadrant, with the goal of aligning engineering efforts to the strategic objectives set by the CSCS management board. This process is structured into several phases that repeat each quarter.

Approximately one month before the end of each quarter, the strategic goals for the upcoming quarter are defined and communicated to the WS. During this phase, CSCS management and architects hold multiple sessions to translate these strategic objectives into high-level engineering initiatives and to prioritize them accordingly. These initiatives are then further refined collaboratively between architects and the respective WS teams.
Each WS develops its local plan by taking into account team member availability, ongoing commitments such as lifecycle management and ticket handling, and the newly prioritized strategic goals.
At the end of the quarter, all engineers, CSCS management, and key partner representatives gather for a two-day planning event. This step consolidates the local WS plans into a cohesive global roadmap for the engineering quadrant. During this event, proposed plans are presented, risks are discussed, and the overall plan is finalized through a vote of confidence involving all participants.
Throughout the quarter, progress is tracked via monthly demos, which showcase outcomes and assess alignment with goals. In addition, biweekly synchronization meetings between WSs help maintain momentum and address cross-cutting concerns.
This process follows the principles outlined in the Scaled Agile Framework (SAFe)~\cite{safe}, and supports agile delivery of complex, multi-platform systems within a service-oriented architecture.

\section{Scientific platforms on Alps}

The Alps infrastructure serves as a research platform for diverse scientific communities, offering robust multi-tenant capabilities. Tenants are categorized based on their level of interaction with the infrastructure. The first category manages platforms and their associated services, seamlessly integrating these with the broader suite of services. The second category operates at the Infrastructure level, independently managing resources with privilege access and overseeing the services deployed on top.

\subsection{MeteoSwiss, ICON-22 platform}

MeteoSwiss, Switzerland's national meteorological agency, operates its numerical weather prediction (NWP) system on the Alps infrastructure~\cite{maurocug25}. MeteoSwiss manages the NWP services while CSCS provides the dedicated platform with specific resources, adhering to strict service-level agreements for high availability, time-critical performance, and reliability. The platform comprises two vClusters: one for production workflows and another for research and development, in total utilizing approximately 100 A100 nodes. The vCluster technology also enables seamless replication and recovery of the platform on alternative infrastructures, ensuring service continuity during disruptions or maintenance.

\subsection{PSI - Merlin7 platform}

The Paul Scherrer Institute (PSI) operates advanced research infrastructures, including cutting-edge light, neutron, and muon sources. These facilities rely on specialized beamlines with demanding data acquisition and processing pipelines, requiring substantial computational resources. As equipment evolves, the growing data processing needs have surpassed the capacity of PSI on-premise HPC clusters. To address this, PSI has begun migrating its clusters to the Alps infrastructure, starting with a pilot platform.  

To maintain autonomy, PSI manages its users, aligns with beamline schedules, and configures specific tools independently. Functioning as an infrastructure-level tenant within Alps, PSI has full control over its allocated nodes, including configuration and reboot capabilities. Its network seamlessly integrates with the segregated Alps platform, enabling it to provide services akin to an on-premise cluster. This approach exemplifies an IaaS model, offering flexibility not typically possible with traditional HPC systems~\cite{HPCIaaS}.

\subsection{User lab - HPC platform}
CSCS provides large-scale HPC resources to national and international research groups, as well as academic institutions, through a competitive proposal submission process. Managed directly by CSCS, the HPC platform offers HPC services while integrating key components of the Alps infrastructure, including approximately 600 GH200 nodes and multicore Rome nodes. This platform serves as an evolution from CSCS's flagship system, Piz Daint, into the Alps infrastructure.


\subsection{Swiss AI - ML platform}
A notable addition to the Alps infrastructure is a dedicated Machine Learning (ML) platform managed by the Swiss AI initiative. What differentiates this platform is its specialized software stack and services, designed specifically for ML workloads, diverging significantly from traditional HPC offerings~\cite{Schuppli25}. The platform leverages containerized execution environments, preconfigured with the ML Python ecosystem, including frameworks like PyTorch. This approach simplifies the deployment and execution of ML tasks, providing an intuitive and efficient environment for ML engineers. 

To meet the computational demands of training advanced AI models, such as large language models (LLMs), this platform allocates a significant portion of Alps resources—approximately 1,300 GH200 nodes—ensuring robust performance and scalability for cutting-edge AI research and long-lasting jobs.

\subsection{EXCLAIM - C\&W platform}
Alps hosts a dedicated climate and weather simulation platform that mirrors the structure of the primary HPC platform and leverages approximately 500 GH200 nodes and 60 PB of storage. This platform supports the objectives of the EXCLAIM project~\cite{exclaim}, which focuses on executing ICON-based climate simulations at kilometer-scale resolution, both regionally and globally. To facilitate scientific autonomy and rapid iteration, EXCLAIM researchers can independently manage and build different versions of the ICON model using the uenv tool.

\subsection{Other platforms}

Currently under development, CSCS is engineering several new platforms to cater to diverse scientific domains. These include a platform for the Worldwide LHC Computing Grid (WLCG), leveraging Grid-specific frameworks; a platform for the Cherenkov Telescope Array (CTA) and another platform for the Square Kilometre Array (SKA) both utilizing Kubernetes for orchestration; and a platform for Materials Cloud, designed to meet specialized requirements such as high-throughput computing (HTC) and the integration of Grid services. 

\begin{figure}[ht]
\centering
\includegraphics[width=\linewidth]{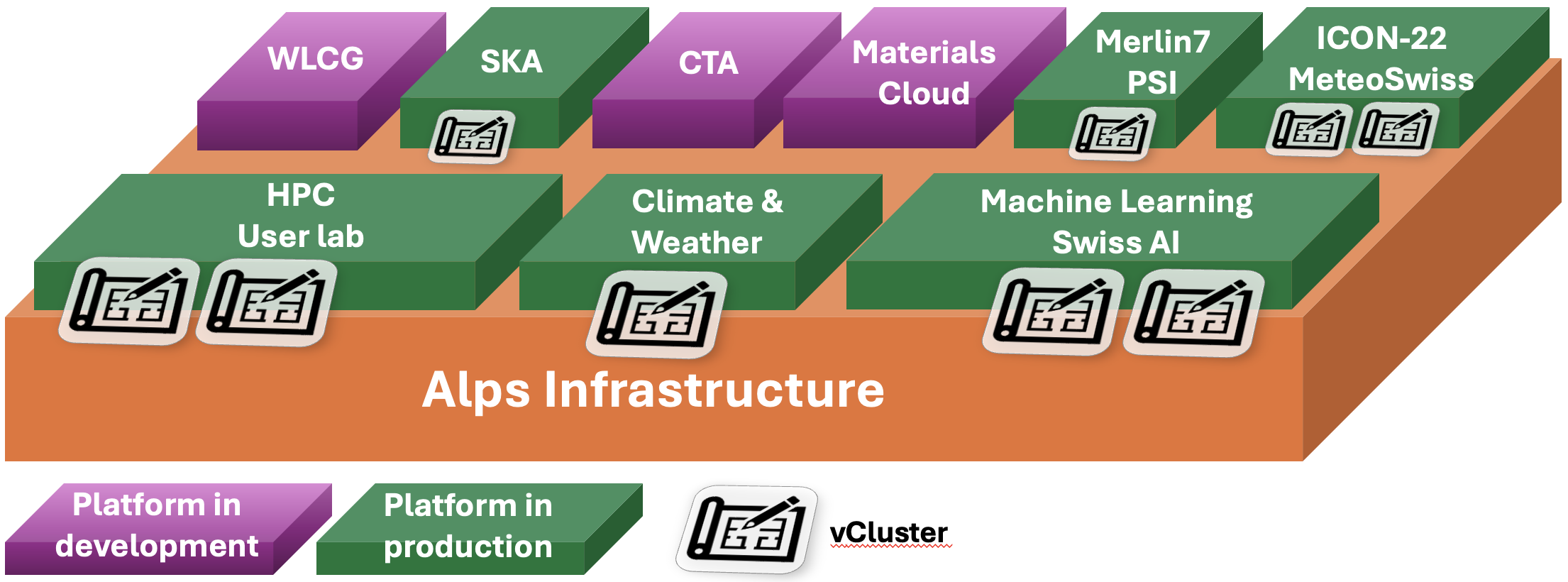}
\caption{Example of platforms running on the Alps infrastructure.}
\label{Alps}
\Description{Diagram showing the platforms available on Alps.}
\end{figure}

Figure~\ref{Alps} presents the set of production and development platforms running on Alps. Each platform is composed of one or multiple vClusters and can offer multiple user environments. CSCS also leverages vClusters for its internal development and technology exploration efforts, utilizing on-demand and temporary vClusters that are not publicly documented or explicitly mentioned.

\section{Discussion}

The vCluster technology and multi-platform setup have been operational on Alps for over a year. In this section, we will explore the key aspects of this journey.

\subsection{Versatility vs. Complexity}

To deliver versatility on the HPC system, CSCS embarked on the development of the vCluster technology, a multi-year effort involving architectural design, technology selection, and iterative solution finding. In parallel, the organization underwent a significant internal restructuring to support a service-oriented architecture capable of managing multiple, diverse platforms. This transformation has been both ambitious and intense. Translating the architectural shift into practice required the creation of new processes, roles, and responsibilities, as well as a cultural change in how services are delivered—through automation, immutable infrastructure, and end-to-end responsibility embedded within each team. These changes, while essential, have posed challenges for many CSCS members, particularly in navigating the increased organizational and technical complexity.

Supporting a heterogeneous landscape of platforms inherently increases the cognitive load for both engineers and users: documentation expands in scope, testing becomes more intricate, and support paths must account for a wider variety of configurations. Teams often face context-switching between different platform configurations and objectives. Moreover, the coordination overhead grows as demonstrated by the quarterly planning.

Another consequence of this versatility is the cost of platform ownership: smaller or community-specific vClusters can be expensive to maintain unless there is reuse across domains or automation reaches a sufficient level of maturity. These trade-offs have prompted ongoing standardization efforts across vClusters, aiming to reduce divergence in configuration and operations while still enabling the flexibility required by scientific users.
Managing Alps today, as described in this paper, requires a high-level of abstraction to grasp the overall system architecture. Meanwhile, its day-to-day operation depends on effective collaboration, role delegation, and trust among teams to maintain cohesion and deliver reliable, secure platforms at scale.

\subsection{Operational experience and early lessons}

Adopting immutable infrastructure principles, and transitioning from traditional system administration to a fully GitOps-based operational model, represents a major cultural and technical shift. The role of a system engineer now closely resembles that of a software developer, requiring a strong understanding of automation pipelines, version control, declarative infrastructure, and release management. This transition has proven challenging for many, requiring significant training, mindset changes, and support.

At the same time, Alps is the first production system to introduce GH200 at scale while also pioneering the vCluster-based versatility model. These twin innovations compound uncertainty: operational behavior of the hardware may still evolve with maturing firmware or drivers, while the surrounding orchestration stack must be actively adapted to support new use cases. In this context, interactions with vendor support have also grown more complex, as responsibilities are split between the infrastructure, service management, and user workloads layers, making complex problem isolation harder to achieve. Anecdotally, the term "operations" has taken on a negative connotation, often associated with firefighting or unclear ownership. Internally, there's a growing preference to use the term "lifecycle management" to describe planned and transparent changes to the system together with "incident management" for unexpected issues, both reflecting a more modern and proactive approach.

While automation significantly reduces manual error and accelerates time-to-delivery, it brings with it a hidden operational cost: debugging failures at runtime becomes more difficult due to the complexity of layered abstractions and distributed logs. Engineers must adhere strictly to software engineering practices—including CI/CD, version control, and test-driven development—to maintain traceability, reproducibility, and observability of the deployed environments. Without this rigor, the debugging process can easily lose track of the actual state of deployed images or services, especially when multiple teams are involved.

\subsection{Automation: promise vs. practice}

Automation has largely delivered on its promise within the Alps system. The use of declarative configurations, version-controlled infrastructure, and merge request-based workflows has significantly accelerated the delivery of services. With reproducibility at the core, changes to vServices are now traceable, auditable, and aligned with modern software engineering practices. The use of release notes for vServices further contributes to operational transparency and accountability.

One of the most notable achievements is the ability to perform rolling updates, e.g., deploying new services or updates without taking the HPC system offline or blocking user access. This marks a significant departure from traditional HPC operations, where system-wide downtimes were a necessary part of the upgrade cycle. In this new paradigm, the ultimate goal of zero-downtime updates is no longer theoretical but increasingly within reach. However, as discussed earlier, when automation workflows fail, the difficulty of troubleshooting increases. This is partly due to the layered complexity and partly due to limited operational maturity: teams are still building the necessary experience, mental models, and observability tooling to quickly diagnose failures across infrastructure, service orchestration, and application layers. Nevertheless, reliable rollback capabilities are critical, the ability to quickly return to a known-good state minimizes disruption and operational risk.

Another powerful outcome of automation is the democratization of contribution. Engineers no longer need privileged system access or manual intervention to test ideas. Thanks to integration pipelines and well-defined testing environments, engineers from various teams can prototype, validate, and deploy changes safely. This has enabled non-system engineers, for example developers or domain scientists, to contribute meaningfully to the platform's evolution and resiliency.

\subsection{Sustainability of the vCluster technology}

The vCluster model offers excellent reproducibility and modularity. However, as the number of platforms and services increases, so does the complexity of the code that needs to be maintained and tested. The sustainability of this model relies on continuous investment in automation, testing, and processes to prevent drift and entropy. The system itself evolves into a large software entity, facing challenges such as software refactoring and technical debt.

One effective solution, alongside good software practices, is to establish a community of multiple centers utilizing the vCluster technology. This community can collaboratively address technical debt and enhance the technology with innovative ideas and solutions. CSCS is committed to providing the complete vCluster technology as open-source software and forming a consortium around it.

\section{Future Directions}

The modular design of the vCluster layers allows for independent enhancements and targeted improvements. For the provisioning infrastructure layer, we plan to transition from CSM to OpenCHAMI while standardizing on the Manta interface for resource management. 
At the service layer, following IBM's acquisition of HashiCorp, we are exploring alternative cloud-native tools for service deployment, such as Crossplane, to enhance flexibility and scalability. Additionally, new services are being introduced, including an HTC scheduler tailored for the Material Clouds platform.

In the long term, our focus is on consolidating the vCluster design to ensure its robustness and adaptability. We aim to collaborate with other infrastructures, such as the Isambard-AI~\cite{IsambardAI} system at the University of Bristol, to expand the adoption and applicability of the vCluster model.

\begin{acks}
CSCS acknowledges the invaluable contributions of our partners and former CSCS team members, in particular Sadaf Alam, whose efforts have significantly shaped the development of the Alps infrastructure.

An AI-generated tool based on ChatGPT built on GPT-4 architecture has been used to enhance the readability of all sections of this document. 
The authors have carefully integrated the AI-suggested edits to preserve the intended meaning.
The AI tool was not used to generate ideas or data. 
\end{acks}


\printbibliography

\end{document}